
\magnification1200
\hsize 130mm
\def\sqr#1#2{{\vcenter{\vbox{\hrule height.#2pt
\hbox{\vrule width.#2pt height#1pt \kern#1pt
               \vrule width.#2pt} \hrule height.#2pt}}}}

\def\autoeq{ {\global\advance\count90 by 1} \eqno(\the\count90) }
\def\autoref{ {\global\advance\refno by 1} \kern -5pt [\the\refno]\kern 2pt}
\def\setup{\count90=0 \count80=0 \count91=0 \count85=0
\countdef\refno=80 \countdef\secno=85 \countdef\equno=90
\countdef\ceistno=91 }
\def\e{\hbox{e}}
\def\Box{\vbox{\hrule
 \hbox{\vrule height 6pt \kern 6pt \vrule height 6pt}\hrule}\kern 2pt}
\input epsf
\setup
\overfullrule=0pt
\centerline{  }
\line{\hfill \hbox{DIAS-STP-95-35 }}
\line{\hfill \hbox{hep-th/9511175}}
\line{\hfill \hbox{November 1995} }
\vskip 2cm
\centerline{\bf Geodesic Renormalisation Group Flow}
\vskip 1.2cm
\centerline{Brian P. Dolan}
\vskip .5cm
\centerline{\it Department of Mathematical Physics,
St. Patrick's College}
\centerline{\it Maynooth, Ireland}
\centerline{and}
\centerline{\it Dublin Institute for Advanced
Studies}
\centerline{\it 10, Burlington Rd., Dublin, Ireland}
\vskip .5cm
\centerline{e-mail: bdolan@thphys.may.ie}
\vskip 1.5cm
\centerline{\bf ABSTRACT}
\bigskip
\noindent
It is shown that the renormalisation group flow in coupling
constant space can be interpreted in terms of a dynamical
equation for the couplings analogous to viscous fluid flow
under the action of a potential.  For free scalar field
theory the flow is geodesic in two dimensions, while for
$D \neq 2$ it is only geodesic in certain limits, e.g.
for vanishing external source. For the 1-D Ising model
the renormalisation flow is geodesic when the external
magnetic
field vanishes.

\vfill\eject
\noindent
{\bf Introduction} \hfill
\bigskip\noindent
The machinery of the renormalisation group has evolved
steadily over the 40 years since its inception and
recently the notion of a geometry on the space of
couplings has received some attention
\autoref\newcount\Sonoda\Sonoda=\refno
\autoref\newcount\Denjoe\Denjoe=\refno,
(for a review in a thermodynamic context see
\autoref\newcount\Ruppeiner\Ruppeiner=\refno).
The aim of this paper is to use these geometrical
notions to interpret the $\beta$-functions as a
dynamical flow and search for a dynamical equation
governing the evolution of the couplings of a quantum
field theory under changes of scale.  It will be argued
that, with a specific choice of metric on the space of
couplings, the flow of the couplings is analogous to
fluid flow in a curved space, under the influence of a
potential, including friction.  Under certain circumstances
the flow is geodesic, with the
constraint that the kinetic energy equals the potential energy.
\bigskip
\noindent
For example for free massive field theories coupled to a
constant source, the mass $m^2$ and the source $j$
parameterise a two dimensional space and it will be
shown that the renormalisation group flow is geodesic
along the $j$-axis $(m^2=0)$ and along the $m^2$-axis
$(j=0)$, but not otherwise, (except in two dimensions
which is special in that all trajectories are geodesics).
\bigskip
\noindent
The concept of geodesic flow requires introducing a
metric on the space of couplings [\the\Denjoe]
\autoref\newcount\Zamolodchikov\Zamolodchikov=\refno
\autoref\newcount\HughIan\HughIan=\refno.
The significance of a metric was highlighted by Zamolodchikov
who utilised the concept in the proof of the c-theorem in
two dimensional field theory [\the\Zamolodchikov].  A
related metric in $D$-dimensions was considered by O'Connor
and Stephens in [\the\Denjoe]
and this is the metric which will be studied here.
\bigskip
\noindent
The construction used by O'Connor and Stephens was to
consider the partition function $Z$ and free energy $W$
to be differentiable functions on the $n$-dimensional
space of couplings, parameterised by $g^a, a=1, \dots, n$.

\noindent
Then from the identity
$$
1\quad = \quad \int D\varphi e^{-S[\varphi]+W}
\autoeq
$$
follows the formula
$$
dW = \langle dS \rangle
\autoeq
$$
where $dS = (\int \Phi_a(x)d^Dx)dg^a$ is an operator
value one - form.  O'Connor and Stephen's metric is
$$
\langle (dS - dW) \otimes (dS - dW)\rangle .
\autoeq
$$
\noindent
In order to be able to pass to the infinite volume limit
densities will be used here. Defining
$$
\tilde\Phi_a(x) \quad = \quad \Phi_a(x) -
\langle\Phi_a(x)\rangle \quad ,
\autoeq $$
we shall take
$$
G_{ab} = \int \langle\tilde\Phi_a(x) \tilde\Phi_b(0)\rangle
d^Dx
\autoeq $$
\newcount\PhiGdef\PhiGdef=\equno
to be the metric.
\bigskip
\noindent
In special co-ordinates in which the action is linear in
the couplings, $G_{ab}$ can easily be derived from the free
energy density $w$, defined via  $W=\int w d^Dx$, from
$$
G_{ab} = -\partial_a\partial_bw.
\autoeq $$
\newcount\Gdef\Gdef=\equno
\noindent
Note however that equation (\the\Gdef) is not covariant under
general co-ordinate transformations, whereas equation
(\the\PhiGdef) is.  In general there may be singularities
in  $\langle\tilde\Phi_a(x) \tilde\Phi_b(y)\rangle$ as
$|x-y|\rightarrow 0$ which are strong enough to render
the integral in (\the\PhiGdef) divergent and it will be
assumed that these can be regularised. It will also be
assumed that
$\langle\tilde\Phi_a(x) \tilde\Phi_b(y)\rangle$ falls of
fast enough for large $|x-y|$ for (\the\PhiGdef) to be finite.
The formalism will now be illustrated with two examples - free
field theory and then the 1-D Ising model.
\bigskip
\noindent
{\bf Free Field Theory} \hfill
\bigskip\noindent
Consider free field theory in $D$ Euclidean dimensions,
with action
$$
S(\varphi;j,m^2) \quad =
\quad \int d^Dx\biggl\{ {1\over 2}\varphi
(- \Box^2+m^2)\varphi+j\varphi\biggr\} .
\autoeq
$$
\bigskip
\noindent
There are two parameters in this theory, $j$ (which
shall be taken to be independent of position for simplicity)
and $m^2.$
The partition function is
$$
Z(j, m^2)=\int D\varphi e^{-S(\varphi;\ j,m^2)}.
\autoeq
$$
\bigskip
\noindent
Performing the Gaussian functional integral gives the
free energy density, using dimensional continuation with
dimension $D$,
$$
w(j,m^2) = - \biggl( {m^2 \over 4\pi}\biggr)^{D/2}
{{{\Gamma(2-{D\over 2})} \over {D({D\over 2}-1)}}} -
{j^2\over 2m^2}.
\autoeq
$$
\noindent
Since the action is linear in $j$ and $m^2$ the metric as
given by (\the\Gdef) is easily calculated giving
$$
ds^2 = - (\partial_a\partial_bw)dx^adx^b = {1\over {m^2}}
\,dj^2 + \left[{{{\Gamma(2-{D\over 2})}
\over 2{(4\pi)^{D\over 2}}}} \,(m^2)^{{D\over 2}-2} +
{j^2\over m^6}\right](dm^2)^2 - {2j\over{m^4}} \, djdm^2.
\autoeq
$$
\bigskip
\noindent
As pointed out in [\the\Denjoe], it is convenient to
use $\phi = \langle \varphi \rangle = - j/m^2$, rather
than $j$ itself as a co-ordinate, because this diagonalises
the metric.  This leads to
$$
ds^2 = m^2 d\phi^2 + {1\over2} \,\, {1\over{{(4\pi)}}^{D/2}}
(m^2)^{{D\over 2}-2} \,\,{\Gamma\biggl(2-{D\over 2}\biggr)}
(dm^2)^2.
\autoeq
$$
\noindent
To expose the geometry more clearly a further change of
co-ordinates is useful.

\noindent
Let $$\theta = 2\sqrt{\pi} \biggl\{{D\over 4} \sqrt{2\over
\Gamma (2-{D\over 2})}\biggr\}^{2\over D} \quad \phi
$$
$$r = {4\over D} \sqrt{\Gamma (2-{D\over 2})\over 2} \,\,
\left({m^2\over 4\pi}\right)^{D\over 4}
\autoeq
$$
\newcount\rdef\rdef=\equno
then
$$
ds^2 = dr^2 + r^{4\over D} d\theta^2.
\autoeq
$$
\newcount\ds\ds=\equno
\noindent
This co-ordinate transformation is singular for $D = 4$, but
perfectly regular for \hbox{$0<D<4$}.
It is clear from (\the\ds) that
$D = 2$ gives the flat metric in $r -\theta$ space, but all
other dimensions lead to a curved  metrics.  The curvature
scalar is
$$
{\cal R} = - {{2(2-D)}\over{D^2r^2}} = - {{(2-D)}\over 4} \,\,
{1\over{\Gamma(2-{D\over 2})}}
\quad \biggl( {{m^2}\over {4\pi}}\biggr)^{-{D\over 2}}
\autoeq
$$
\newcount\GaussRicci\GaussRicci=\equno
which is equivalent to the large volume limit of [\the\Denjoe].
\bigskip
\noindent
Note that ${\cal R} = 0$ in $D = 4$ for finite $m^2$, though
it is positive for finite $r$, due to the singular nature of
the co-ordinate transformation (\the\rdef) in four dimensions.
For $D=0$,
${\cal R}=-(1/2)$ for finite $m^2$, giving the Lobachevski plane
for the Gaussian distribution in ordinary statistics,
\autoref
\newcount\Amari\Amari=\refno. It is, of course, $m^2$ which
is the physical parameter.
\bigskip
\noindent
We shall now turn to a discussion of the renormalisation flow
in this geometry. Since the model is a free field theory, the
couplings simply scale according to their canonical dimensions,
and $\beta$-functions can be defined by using dimensionless
parameters
$$
m^2\rightarrow m^2/\kappa^2, \quad \quad \phi \rightarrow
\phi/\kappa^{{D\over 2}-1},
$$
where $\kappa$ is a fiducial scale (renormalisation point),
giving $$
\beta ^{m^2} = \kappa {{dm^2}\over{d\kappa}} = - 2m^2
$$

$$
\beta^\phi = \kappa {{d\phi}\over {d\kappa}} =
- \biggl({D\over 2}-1\biggr)\phi
\autoeq
$$
\noindent
or $$
\beta^r = - {D\over 2}r \quad \quad \beta^\theta =
- \biggl( {D\over 2}-1\biggr)\theta
\autoeq
$$
\newcount\betadef\betadef=\equno
in the $(r-\theta)$ co-ordinate system.  The vector field
$\vec\beta = - ({D\over 2}-1)\theta {{\partial}\over
{\partial\theta}} - {D\over 2} r {{\partial}\over
{\partial r}}$ is plotted in figures 1, 2 and 3 for
dimensions 1, 2 and 3 respectively.
The two
dimensional case is plotted in polar co-ordinates in figure 2
since it is
clearly natural to make $\theta$ periodic in this instance,
with period $2\pi$.
This is reflected in two dimensional conformal field theory
where the field $\vartheta = \sqrt{2\pi}\varphi$ is not a
primary field, but the operator $:\e^{ie\vartheta}:$ is
primary, with correlator
$$<:\e^{ie\vartheta(x)}::\e^{ie\vartheta(y)}:>
 \approx {1\over\vert x-y\vert^{e^2}},\autoeq$$
\autoref\newcount\Cardy\Cardy=\refno. It appears, therefore,
that the operator $z={m\over\sqrt{2\pi}}\e^{i\vartheta}$
is the most natural, not only from the point of view of
conformal field theory but also from the geometrical point
of view taken here.
There is no
obvious reason for making the fields periodic in the other
dimensions.
\bigskip
\noindent
It seems natural to enquire if the vector flows plotted
in the figures are related in any way to the metric in
equation (\the\ds).  Clearly the integral curves of
(\the\betadef) in $D = 2$ are radial and thus geodesics
of the metric (\the\ds).  This is not true for $D\neq 2$.
If a curve $x^a(t)$ is parameterised by $t$ and has tangent
vector $\zeta^a = {{dx^a}\over {dt}}$ the condition that
$x^a(t)$ be a geodesic is given by the geodesic equation
$$
\zeta^b \nabla_b\zeta^a = \lambda \zeta^a
\autoeq
$$
\newcount\GeoEq\GeoEq=\equno
where the right hand side, $\lambda \neq 0$, allows for the
possibility that $t$ might not be affine
\autoref\newcount\Hawking\Hawking=\refno.
\bigskip
\noindent
If $\vec \beta$ in (\the\betadef) is used in the geodesic
equation
(\the\GeoEq) with the metric (\the\ds) one finds the
following conditions on $r(t)$ and $\theta (t)$
$$
\biggl( {D\over 2}\biggr)^3 r -
\biggl({D\over 2}-1\biggr)^2  r^{{4-D}\over D}\theta^2 =
- \biggl({D\over 2}\biggr)^2 \lambda r
$$
$$
\biggl({{D^2}\over {4}}-1\biggr) \theta =
- \biggl({{D}\over {2}} -1\biggr)\lambda\theta.
\autoeq
$$
\noindent
If $D\neq 2$, the only solutions, for $0<D<4$, are either $r =0$
(i.e. $m^2=0$) with $\lambda = - \bigl({{D}\over {2}}
+ 1\bigr)$ or $\theta = 0$ with $\lambda = - {{D} \over
{2}}$.  Thus the bold lines in figures 1 and 3 are geodesics.
The other curves fail to satisfy the geodesic condition
(\the\GeoEq) for $D\neq 2$, and can be interpreted as
being repulsed from the $\phi$-axis by a \lq\lq force".
This conclusion also holds for $D=4$ where the only
renormalisation group flows
which are geodesic are $m^2=0$ with $\lambda=-3$ or $\phi=0$
with $\lambda=-2$.
\bigskip
\noindent
{\bf The 1-D Ising Model} \hfil
\bigskip\noindent
Clearly it is of interest to extend the above analysis beyond free
field theory to the case
of a non-trivial interacting theory. Consider therefore
one of the simplest interacting theories - the one dimensional
Ising model.

\noindent The one dimensional Ising model on a periodic
lattice of N sites is described in
\autoref\newcount\Baxter\Baxter=\refno.
The partition function is
$$
Z_N=\sum_{\{\sigma\}} exp \left[ {K \sum_{j=1}^N \sigma_j
\sigma_{j+1} + h
\sum_{j=1}^N \sigma_j} \right]
\autoeq
$$

\noindent where $K = {{J} \over {kT}}$ and $h = {{H} \over
{kT}}$, with J the spin coupling
and H the external magnetic field, (periodic boundary
conditions require
$
\sigma_{N+1} \equiv \sigma_1)
$.
$Z_N(K,h)$
can be conveniently expressed in terms of the transfer matrix
$$
V=\pmatrix{
V_{++}&V_{+-}\cr
V_{-+}&V_{--}\cr}
\quad = \quad
\pmatrix{
e^{K+h}&e^{-K}\cr
e^{-K}&e^{K-h}\cr
} \autoeq
$$

\noindent as $Z_N = TrV^N$.
\bigskip
\noindent Diagonalising $V$ gives the eigenvalues
$$
\lambda_\pm=e^K\left\{\cosh h\pm\sqrt{\sinh^2 h+e^{-4K}}\right\}.
\autoeq
$$
\newcount\Eigenvalues\Eigenvalues=\equno
Thus
$$
Z_N=\lambda_+^N\left[{1+}\biggl({\lambda_-
\over \lambda_+}\biggr)^N\right]
\autoeq
$$
and in the limit of large $N$
$$ (1/N)\ln Z_N = K + \ln\left(\cosh h + \sqrt{\sinh^2 h+
e^{-4K}}\right),$$
which is essentially the negative of the free energy per unit
volume.
\bigskip
\noindent
Using equation (\the\Gdef) the line element is easily obtained
$$\eqalign{ ds^2&=
  {\e^{-4K}\over\bigl(\sinh^2h+\e^{-4K}\bigr)^{3/2}}\cr
      \times & \left[
         {4\e^{-4K}\cosh h + 8\sinh^2 h
             \bigl(\cosh h+\sqrt{\sinh^2h+\e^{-4K}}\bigr)
         \over{\bigl(\cosh h+\sqrt{\sinh^2h+\e^{-4K}}
               \bigr)^2}}\ dK^2\right. \cr
 	&  \left. {\vrule height16pt width0pt}\qquad\qquad
        + 4\sinh h \ dKdh + \cosh h \ dh^2\right]
      .\cr}
\autoeq $$
\newcount\ChaosG\ChaosG=\equno
It is convenient to change to a set of co-ordinates in
which the metric is diagonal. To this end define
$\rho=\e^{2K}\sinh h$ and use $K$ and $\rho$ as
co-ordinates. The metric is then
$$
ds^2={1\over\sqrt{(1+\rho^2)(\e^{4K} +\rho^2)}}
       \left[
   {4\;\e^{4K} dK^2\over
       \bigl(\sqrt{1+\rho^2}+\sqrt{\e^{4K}+\rho^2}\bigr)^2}
	 + {d\rho^2\over (1+\rho^2)}\right].
\autoeq
$$
The Ricci scalar is easily calculated and is found to be
$$
{\cal R}={1\over 2}\left[
     1+\sqrt{\e^{4K} +\rho^2 \over 1+\rho^2}\right]
\autoeq
$$
or, in terms of the physical co-ordinates $K$ and $h$,
$$
{\cal R} = \left({1\over 2}\right)
\left( 1 + {\e^{2K}\cosh{h}\over\sqrt{\e^{4K}\sinh{h}^2 + 1}}
      \right).\autoeq
$$
This expression was first obtained by Janyszek and Mruga{\l}a,
\autoref\newcount\JanMru\JanMru=\refno,
and the function is depicted in Fig. 4.
Note that the curvature diverges at the critical point $T=0$,
($K\rightarrow\infty$).

\noindent
To investigate the behaviour of the renormalisation
flow in this lattice model the $\beta$-functions are replaced
by a discrete recursive map,
\autoref\newcount\Fisher\Fisher=\refno.
In its simplest form this map is obtained by asking: can
one find new couplings $K^\prime$ and $h^\prime$ such that
$$
Z_{N\over2}(K^\prime,h^\prime)=A^NZ_N(K,h) \ ?
\autoeq
$$\newcount\part\part=\equno
($A$ is a normalisation factor.)

\bigskip
\noindent Equation (\the\part) is easily satisfied by demanding
$$
\pmatrix{
e^{K^\prime+h^\prime}&e^{-K^\prime}\cr
e^{-K^\prime}&e^{K^\prime-h^\prime}\cr
}
\quad = A^2\quad
\pmatrix{
e^{K+h}&e^{-K}\cr
e^{-K}&e^{K-h}\cr
}^2 \autoeq
$$
giving the recursive formulae:
$$
{e^{2h^\prime}=e^{2h}}\quad {\cosh (2K+h)\over\cosh(2K-h)}
\autoeq
$$\newcount\newconst\newconst=\equno

$$
e^{4K^\prime}=\quad {\cosh(4K)+\cosh(2h)\over 2\cosh^2(h)}.
\autoeq$$\newcount\recrel\recrel=\equno
The normalisation factor $A$ is unimportant for the present
analysis.
\bigskip
\noindent Note that the combination $e^{2K^\prime}\sinh(h^\prime)
 = e^{2K} \sinh(h)$ is a renormalisation
transformation invariant. Strictly speaking this transformation
is not always invertible and so it is incorrect to refer to it
as a renormalisation group transformation.
The renormalisation flow is depicted in Fig. 5.
\bigskip
\noindent
Since the renormalisation transformation is a discrete map
here, the $\beta$-functions cannot be defined in terms of
continous a vector field. However, it still makes sense to
ask if any of the flow curves in Fig. 5 are geodesics of the
metric (\the\ChaosG). The answer is that only the special
curve running between the two fixed points
$K=0$ and $K=\infty$ is a geodesic,
i.e. the seperatrix $h=0$, the other flow lines are
\lq\lq repulsed'' from the $K$-axis and never cross it, but
crowd together asymptotically in the manner
of a caustic
in geometrical optics or fluid mechanics. The geodesic
distance between the two fixed points is $\pi/2$.
This behaviour is suggestive of  a dynamical interpretation
of renormalisation flow.

\bigskip\noindent
{\bf General Considerations} \hfil
\bigskip\noindent
One is led to enquire into the nature of the \lq\lq force''
and to investigate the possibility of a dynamical description
of the renormalisation group flow in terms of an equation
for the $\beta$-functions.
It will be shown that the $\beta$-functions obey a dynamical
equation analogous to the viscous flow of a fluid in a curved
space under the influence of a potential, with a constraint
which dictates that the kinetic energy equals the
potential energy.
\bigskip
\noindent
The argument hinges on the renormalisation group equation
for the two point correlators of the theory
$$
\kappa {{\partial} \over {\partial \kappa}}
\langle \tilde\Phi_a(x)\tilde\Phi_b(y)\rangle +
\beta^c\partial_c \langle\tilde\Phi_a(x)\tilde\Phi_b(y)\rangle
+\partial_a\beta^c \langle\tilde\Phi_c(x)\tilde\Phi_b(y)\rangle
+\partial_b\beta^c \langle\tilde\Phi_a(x)\tilde\Phi_c(y)\rangle
=0.
\autoeq
$$
\newcount\ClumsyRG\ClumsyRG=\equno
\noindent
If the couplings are scaled to be dimensionless, then all
$\tilde\Phi_a(x)$ have canonical mass dimension $D$, thus the
usual scaling argument gives
$$
\bigg(x^\mu {{\partial}\over{\partial x^\mu}} + y^\mu
{{\partial}\over{\partial y^\mu}}\bigg)
\langle\tilde\Phi_a(x)\tilde\Phi_b(y)\rangle + 2D
\langle\tilde\Phi_a(x)\tilde\Phi_b(y)\rangle
$$
$$
=-\beta^c \partial_c\langle\tilde\Phi_a(x)\tilde\Phi_b(y)\rangle
-\partial_a\beta^c \langle\tilde\Phi_c(x)\tilde\Phi_b(y)\rangle
$$
$$
-\,\, \partial_b\beta^c \langle\tilde\Phi_a(x)\tilde\Phi_c(y)
\rangle .
\autoeq
$$
\bigskip
\noindent
Integrating over all $y$ and using translational invariance
leads to
$$
\beta^c \partial_c G_{ab} + \left(\partial_a \beta^c\right) G
_{cb} + \left(\partial_b \beta^c\right)G_{ca}
 =- D G_{ab}.
\autoeq
$$
\newcount\NaiveRG\NaiveRG=\equno
\noindent
This equation states that $\vec\beta$ is a conformal Killing
vector for the metric $G_{ab}$ and it is straightforward to
check this property using (\the\ds) and (\the\betadef).
That the renormalisation group equation can be written in
terms of a Lie derivative was noted by L\"assig in
\autoref\newcount\Lassig\Lassig=\refno
and explored in more detail
in\autoref\newcount\geomrg\geomrg=\refno.
A subtlety arises in $D = 4$ in that the integral over $y$
diverges and $G_{m^2m^2}$ is infinite, due to the short
distance singularity as $y\rightarrow x$.  However, if
$4 - D = \epsilon$ is kept positive and the limit
$\epsilon\rightarrow 0$ is only taken after the scalar
curvature (\the\GaussRicci)
is calculated, it appears that this is a co-ordinate
singularity rather than a pathology in the geometry.
\bigskip
\noindent
If the Lie derivative on the left hand side of equation
(\the\NaiveRG) is written in terms of the Levi-Civita
connection for the metric $G_{ab}$
$$
\beta^c \partial_c G_{ab} + \left(\partial_a \beta^c\right)
G_{cb}+ \left(\partial_b\beta^c\right) G_{ac} =
 \left(\nabla_a\beta^c\right)G_{cb} +
\left(\nabla_b\beta^c\right)G_{ac}
\autoeq
$$
\newcount\geodRG\geodRG=\equno
and then equation (\the\NaiveRG) is contracted with
$\beta^a$, one finds
$$
\beta^b\nabla_b\beta^a = - G^{ab}\nabla_bU - D\beta^a
\autoeq
$$
\newcount\fric\fric=\equno
where the \lq\lq potential" is given by
$U={1\over 2}\beta^b\beta^cG_{bc}$.  The right hand side can be
interpreted as a force, and the potential, $U$, pushes
$\vec\beta$ away from geodesic flow unless
$\nabla_bU \propto \beta_b$. The second term on the
right hand side of (\the\fric) plays the role of
an isotropic frictional force.
\bigskip
\noindent
This dynamical picture is compatible with the analysis of
geodesic flow for free field theories since in this case
$U={1\over 2} \bigl\{ {{D^2r^2}\over{4}} + \bigl( {{D\over 2}
-1}\bigr)^2 \theta^2r^{4/D}\bigr\}$
giving
$$
G^{ab}\partial_bU =
{{\bigl\{ ({D\over 2})^2r +
{2\over D}({D\over 2}-1)^2\theta^2r^{{4\over D} -1} \bigr\} }
\atopwithdelims()
\qquad {\bigl({D\over 2}-1\bigr)^2 \theta \qquad }}
\autoeq $$
and $$G^{ab}\partial_bU = -{D\over 2}\beta^a
\qquad\qquad\hbox{for}\quad \,\,\, \theta = 0$$
$$G^{ab}\partial_bU = -\biggl({D\over 2}-1\biggr)\beta^a
\quad\hbox{for}\,\quad r = 0,\quad D<4 \autoeq $$
so both these cases result in geodesic flow from equation
(\the\fric).
Note the special case of two dimensions where $U=(1/2)r^2$
is the harmonic oscillator potential.
\bigskip
\noindent
The derivation of equation (\the\fric) relied only on the
renormalisation group equation (\the\ClumsyRG) and the fact
that the composite operators, $\tilde\Phi_a$, have canonical
dimension $D$, and so is completely general, even for
interacting field theories.  However, the question of
singularities in $\langle \tilde\Phi_a(x) \tilde\Phi_b(y)\rangle$
as $| x -y |\rightarrow 0$ must be addressed.
If $\langle \tilde\Phi_a(x)\tilde\Phi_b(y)\rangle \sim
{1\over{|x-y|^k}}$ with $k<D$, then $G_{ab} =
\int d^Dx\langle\tilde\Phi_a(x)\tilde\Phi_b(0)\rangle$
is finite (assuming the correlator $\tilde\Phi_a(x)
\tilde\Phi_b(0)\rightarrow 0$ fast enough for large
$|x|$). But if $k\geq D$, then $G_{ab}$ will be singular
and must be regularised. Singular metric components do not
necessarily indicate singular geometry and the curvature
may be finite as the regulator is removed, as the example of
free field theory in $D=4$ shows.
\bigskip
\noindent
In general, however, it may prove necessary to subtract
counter-terms from the two point correlators.  This has the
effect of modifying equation (\the\NaiveRG), so that it reads
$$
\beta^c\partial_cG_{ab} + (\partial_a\beta^c)G_{cb} +
(\partial_b\beta^c)G_{ac} = -DG_{ab} - \chi_{ab}
\autoeq
$$
\newcount\Fric\Fric=\equno
where $\chi_{ab}$ is a new tensor introduced by the
subtraction procedure (see for example [4]).
The dynamical equation (\the\fric) is
then modified by the introduction of a non-isotropic
\lq\lq friction'' and
becomes
$$
\beta^b\nabla_b\beta^a = - G^{ab}\partial_bU - D\beta^a
- {\chi^a}_b\beta^b.
\autoeq
$$
\newcount\DynRG\DynRG=\equno
The condition for geodesic flow is now $G^{ab}\partial_bU
\propto \beta^a$ and ${\chi^a}_b\beta^b \propto \beta^a$.
Equation (\the\DynRG) holds even for renormalisation schemes
in which
the $\beta$ functions have explicit $\kappa$ dependence.
\bigskip

\noindent
In conclusion, it has been demonstrated that the metric
(\the\PhiGdef) can give geodesic renormalisation group flow
in free field theories, under certain circumstances.  This
has been interpreted in terms of the dynamical equation
(\the\fric), analogous to fluid flow with friction in a
curved space, under the influence of a potential with the
constraint $U = {1\over 2}\mid\beta\mid^2$, similar to the
virial theorem for a collection of harmonic oscillators in
statistical equilibrium.
For a general theory the extra non-isotropic frictional term
in (\the\DynRG) may be necessary.
Geodesic renormalisation flow has also been demonstrated
in the one dimensional Ising model for the crossover between
the two fixed points at $T=\infty$ and $T=0$ with vanishing
magnetic field.
\bigskip

\noindent
Clearly it would be of particular interest to examine these
ideas in a non-trivial interacting field theory, where it is
to be expected that the more general dynamical equation
(\the\DynRG) will be relevant and work is in progress on this.
\bigskip
\noindent It is a pleasure to thank Denjoe O'Connor for many
useful discussions Concerningl the renormalisation group.
\vfill\eject
\noindent{\bf References}
\bigskip
\item{[\the\Sonoda]}H. Sonoda, {\sl Connection On The Theory
Space}, UCLA pre-print 93/TEP/21
\item{}{\sl Geometrical Expression For Short-Distance
Singularities In Field Theory},
\item{}UCLA pre-print 94/TEP/20
\item{}{\sl The Energy-Momentum Tensor In Field Theory I},
UCLA preprint 95/TEP/10
\item{}{\sl The Energy-Momentum Tensor In Field Theory II},
UCLA pre-print 95/TEP/31
{\item{[\the\Denjoe]}D. O'Connor and C. R. Stephens
{\sl Geometry, the renormalisation group and gravity} in
{\sl Directions in General Relativity} }
\item{}Proceedings of the 1993 International Symposium
(Maryland), Vol. 1
\item{}eds. B.L. Hu, M.P. Ryan Jr. and C.V. Vishevshawara,
C.U.P. (1993)
\item{[\the\Ruppeiner]}G. Ruppeiner, Rev. Mod. Phys.
{\bf 67}, 605, (1995)
\item{[\the\Zamolodchikov]}A.B. Zamolodchikov, Pis'ma Zh. Eksp.
Teor. Fiz.  {\bf {43}}, 565, (1986)
\item{[\the\HughIan]}I. Jack and H. Osborn, Nucl. Phys.
{\bf{B343}}, 647, (1990)
\item{[\the\Amari]}S. Amari, {\sl Differential Geometric Methods
in Statistics}\hfil\break
Lecture Notes in Statistics {\bf 28}, Springer, (1985)
\item{[\the\Cardy]}J.L. Cardy, {\sl Conformal Invariance
and Statistical Mechanics} \hfil\break
in {\sl Fields, Strings and Critical Phenomena}\hfil\break
Les Houches lectures, Session XLIX 1988, Elsevier (1989)
\item{[\the\Hawking]}S.W. Hawking and G.F.R. Ellis, {\sl
The Large Scale Structure Of Space-Time} \hfil\break
C.U.P., (1973)
\item{[\the\Baxter]}R.J. Baxter, {\sl Exactly Solved Models In
Statistical Mechanics}\hfil\break
Academic Press, (1982)
\item{[\the\JanMru]}H. Janyszek and R. Mruga{\l}a, Phys. Rev.
{\bf A 39}, 6515, (1989)
\item{[\the\Fisher]}D.R. Nelson and M.E. Fisher, Ann. of Phys.
(N.Y.),
{\bf 91}, 226, (1975)
\item{[\the\Lassig]}M. L\"assig, Nucl. Phys.  {\bf{B334}},
652, (1990)
\item{[\the\geomrg]}B.P. Dolan, Int. J. Mod. Phys. {\bf A 9},
1261, (1994)
\vfill\eject
\nopagenumbers
\epsfxsize=125mm
\epsffile{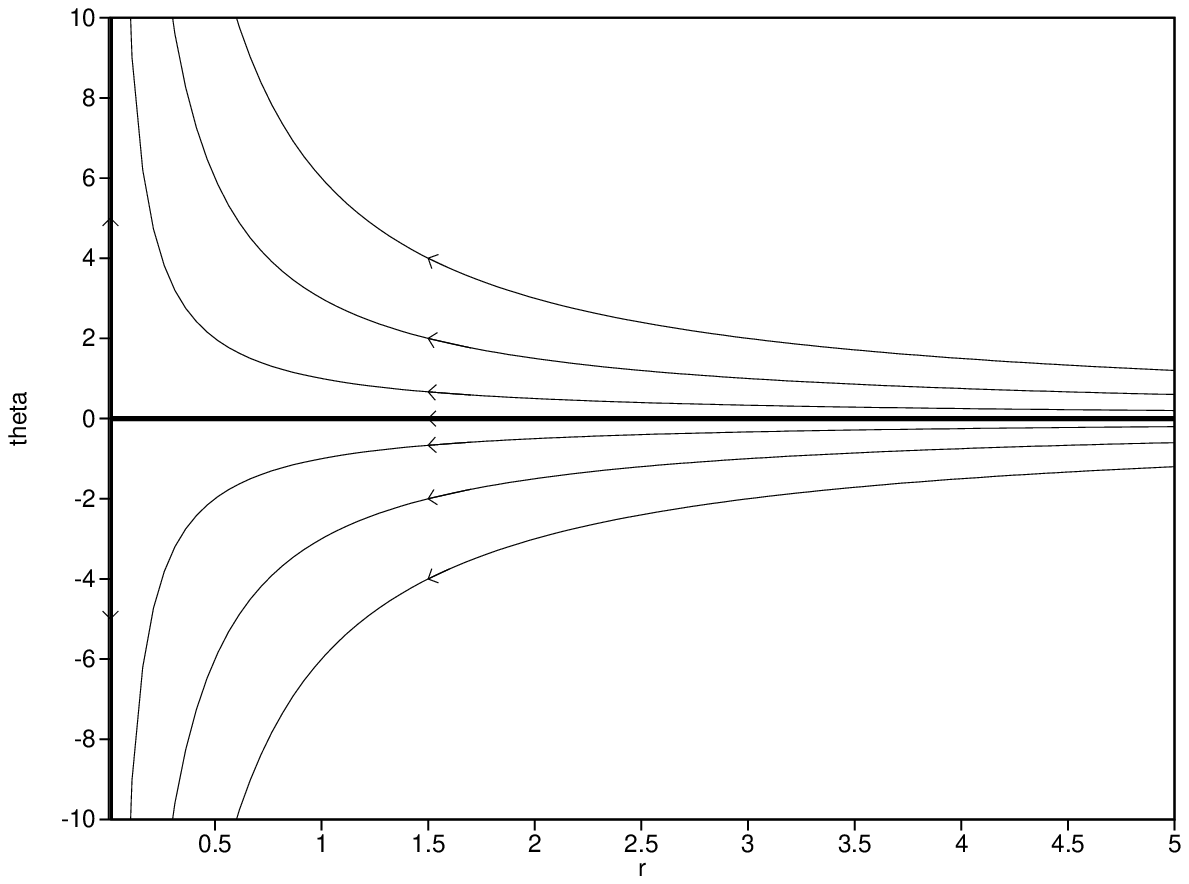}
\centerline{{\bf Fig. 1:} RG flow for free field theory in 1-D}
\centerline{The bold lines are geodesics}
\vfill\eject
\epsffile{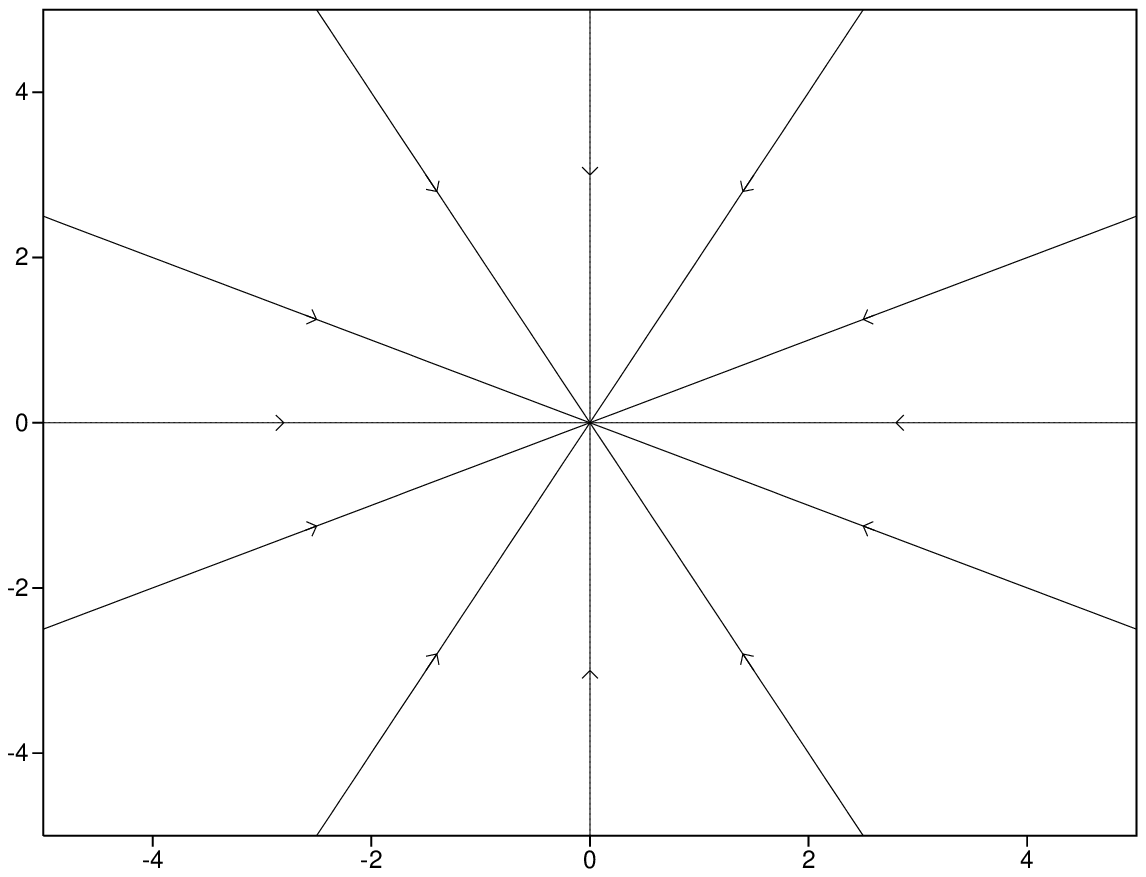}
\centerline{{\bf Fig. 2:}  RG flow for free field theory in 2-D }
\centerline{All RG trajectories are geodesics}
\vfill\eject
\epsffile{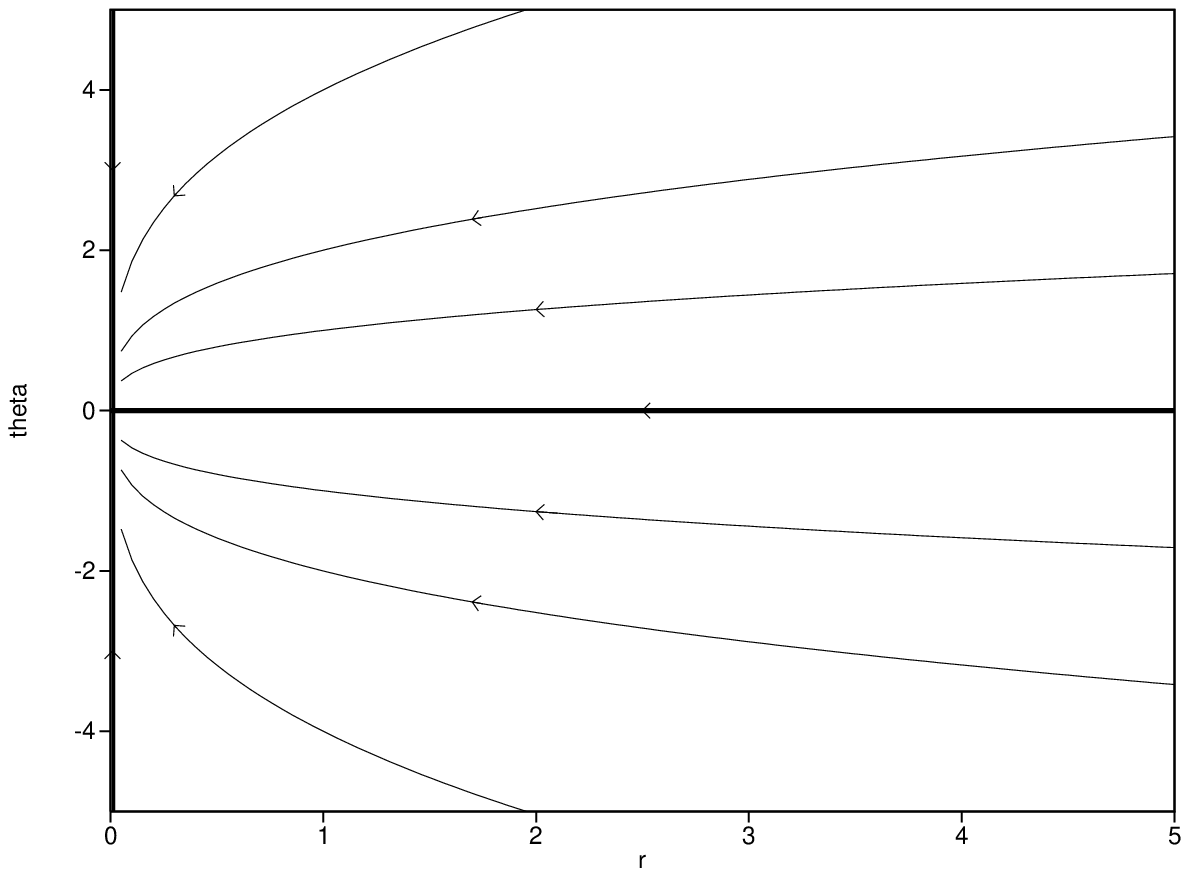}
\centerline{{\bf Fig. 3:} RG flow for free field theory in 3-D}
\centerline{The bold lines are geodesics}
\vfill\eject
\epsfxsize=110mm
\hskip 1truecm
\includegraphics{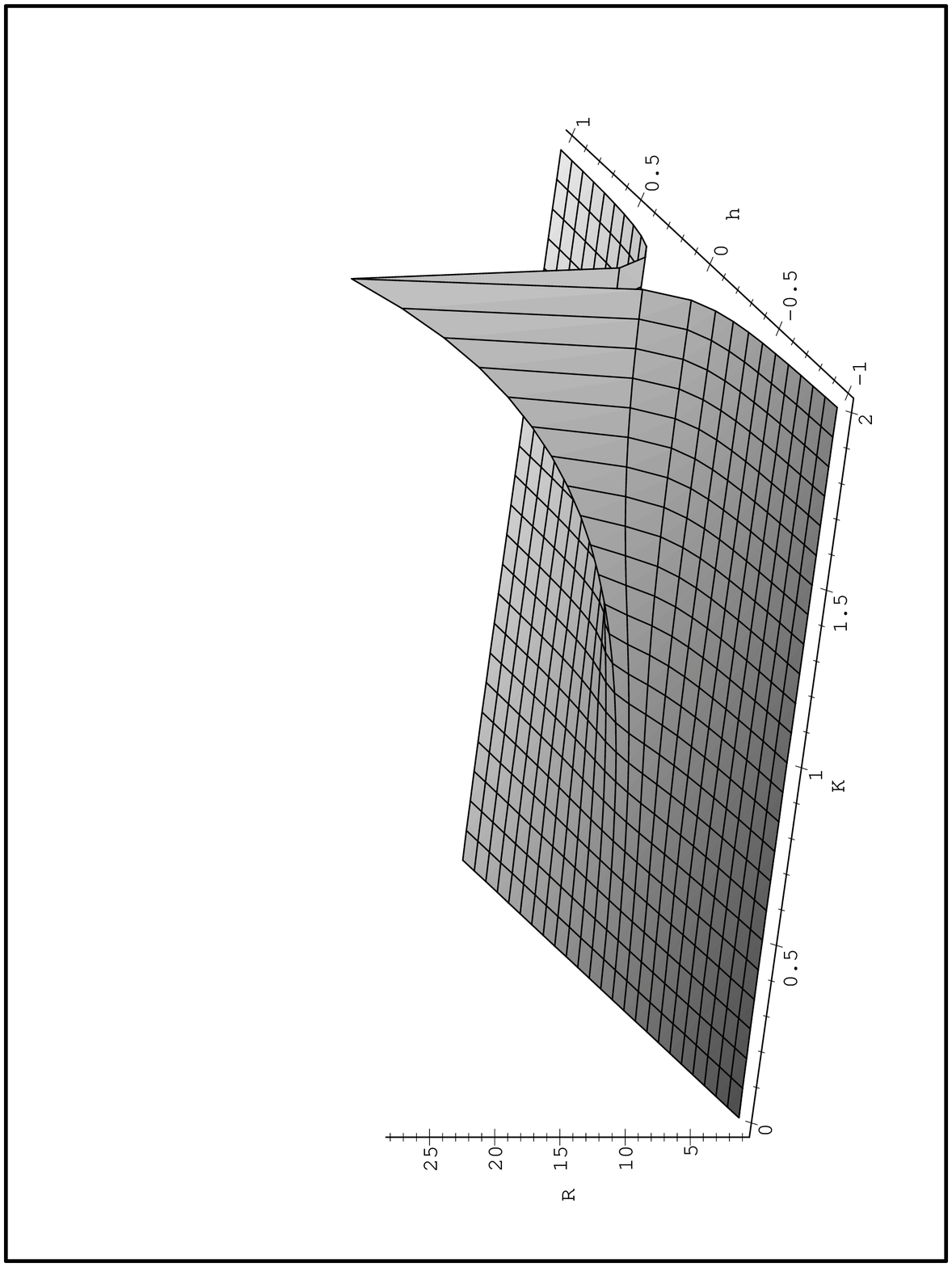}
\vskip 10cm
\centerline{{\bf Fig. 4:} Ricci curvature for the 1-D Ising model}
\centerline{The line of the central ridge is a geodesic}
\vfill\eject
\epsfxsize=125mm
\epsffile{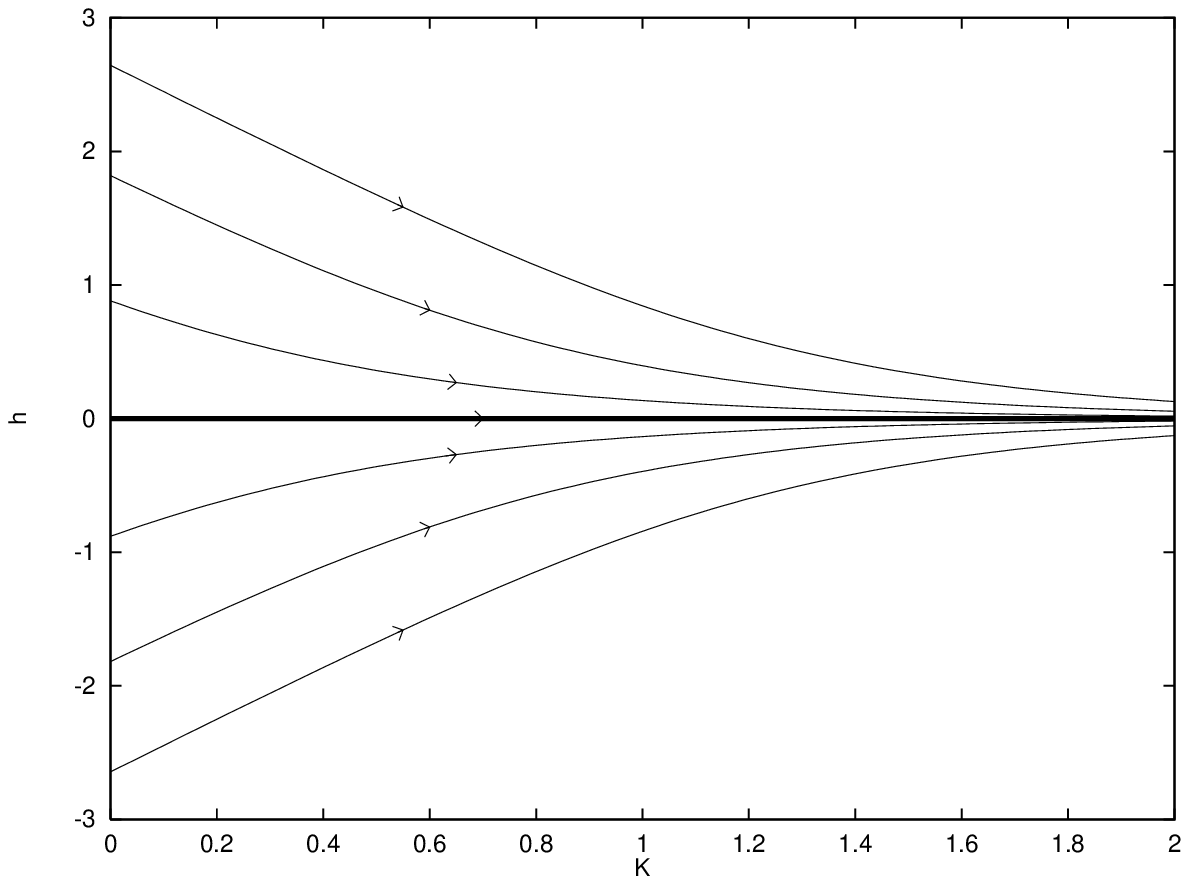}
\centerline{{\bf Fig. 5:} The renormalisation flow for the
1-D Ising model}
\centerline{The $K$-axis ($h=0$) is a geodesic}
\vfill\eject
\end